# A Cost-and-Place-Utility Model of the Move-versus-Commute Decision

**Shawn Berry, DBA[1]***

**June 28, 2026**

[1]William Howard Taft University, Lakewood CO, USA
*Correspondence: shawnpberry@gmail.com

---

## Abstract

Deciding where to live involves a complex balance between commuting and moving, as households must weigh housing affordability, transportation expenses, access to workplaces, and social ties. Traditional urban economic theories focus on the balance between housing expenses and commuting costs, while modern studies also consider housing affordability, transportation access, and utility maximization. However, few studies have combined these elements into a clear mathematical model that can be used for both policy analysis and household decision-making. This paper introduces an algebraic model for deciding whether to commute or move, expanding on traditional residential location theories by including direct housing and commuting expenses, income-related affordability limits, indirect social and service access costs, and location-based utility within a single utility-maximization framework. The model uses the common 30% housing affordability rule as a constraint, acknowledging that residential choices are also shaped by social networks, access to institutions, neighborhood ties, and quality-of-life factors. The decision rule derived from the model integrates direct financial costs with weighted social benefits and indirect access costs to assess when moving offers more overall utility than staying put and commuting. Unlike complex discrete-choice, nested-logit, or agent-based models, this framework offers a mathematically clear, understandable, and flexible decision model that can easily be expanded to include more household characteristics, transportation options, or policy factors. The model advances urban economics, migration studies, and housing affordability research by providing a practical analytical tool for assessing residential mobility decisions within financial and behavioral limits.

## 1. Introduction

The decision to either commute or relocate—choosing between enduring a daily journey to work or moving closer to the workplace—is a crucial aspect of urban migration and labor market dynamics. Traditionally examined from the perspective of utility maximization, this intricate decision involves balancing the direct costs of commuting with the advantages of particular residential areas. While foundational models focus on economic factors, such as housing costs and travel time, recent literature increasingly highlights the role of personal preferences, such as closeness to family, social networks, and urban amenities. This review integrates these theoretical perspectives and examines how housing affordability influences mobility choices, specifically considering the 30% income guidelines of the US

Department of Housing and Urban Development (HUD) as a significant benchmark for individual well-being. The commute-or-move decision is a key issue in urban economic geography because it connects residential location, workplace accessibility, household affordability, transportation expenses, and place attachment within a single decision-making framework. Essentially, a household must decide whether to stay in location A and incur ongoing commuting expenses to a job in location B or move to location B and exchange higher or different housing costs to reduce commuting expenses. However, the existing literature clearly shows that this decision is rarely limited to rent and fuel costs alone. Households assess residential options based on combinations of housing quality, neighborhood characteristics, job accessibility, school and service access, transport availability, income limitations, and social connections (Lee and Waddell, 2010; Lee et al., 2010; Mirzahossein et al., 2023). Thus, the commute-or-move decision is best understood as a constrained utility problem in which direct financial costs interact with indirect time, social, institutional, and psychological costs. This issue has gained importance as metropolitan housing costs have increased relative to income and as job opportunities have become geographically separated from affordable housing. The 30% rent-to-income rule remains a commonly used affordability standard; however, recent studies argue that this threshold is incomplete because it often overlooks transportation, utilities, childcare, healthcare, and other residual-income burdens (Airgood-Obrycki et al., 2022; Gabriel & Painter, 2020; Shamsuddin & Campbell, 2021). In the context of the commute-or-move decision, this limitation is particularly significant because a dwelling that seems affordable based solely on rent may become unaffordable when commuting costs are factored in, whereas a higher-rent residence near employment may be justified if it reduces transportation expenses and time commitments (Ganning, 2017; Makarewicz et al., 2020; Situ & Kim, 2024).

### 1.1 Literature Review

#### 1.1.1 *Theoretical and Classical Models of Residential Location*

Classical urban economic theories lay the groundwork for understanding the direct cost aspect of the commute-or-move model. Alonso's (1964) bid-rent theory and Muth's (1969) housing framework view residential location as a balance between the costs of land or housing and the ease of reaching employment. In this context, households located at a greater distance from employment hubs often benefit from reduced housing prices, yet they face increased commuting expenses., whereas those nearer to employment pay a premium for proximity. Wheaton's (1977) bid-rent perspective also interprets housing demand as a spatial equilibrium issue, where land use and commuting expenses are interrelated. Subsequent research on residential location has expanded beyond classical monocentric assumptions, acknowledging that households do not select locations based solely on a single central business district. Modern metropolitan areas have multiple employment hubs, diverse housing markets, varied household types, and varied accessibility requirements. Discrete-choice and nested-logit models address this complexity by considering residential location as a probabilistic selection among options with varying costs, qualities, and accessibility features (Lee and Waddell, 2010; Mirzahossein et al., 2023; Wang et al., 2021). Lee and Waddell (2010) are particularly pertinent as they view residential mobility and location choice as interconnected phases: households initially decide whether to move or stay, followed by an assessment of potential locations. This two-step reasoning closely aligns with the current commute-or-move model, which first assesses affordability and direct costs, and then evaluates overall utility. The literature also indicates that work accessibility alone is inadequate. Lee et al. (2010) demonstrate that residential choices are shaped by factors such as nonwork accessibility, trip chaining, and household-specific travel needs, even when work-related access is considered. A household relocating to B might still face the costs of maintaining connections with doctors, churches, family, legal services, or associations in A. Conversely, a household remaining in A might incur expenses to access amenities, educational opportunities, healthcare, or career networks concentrated in B. Therefore, the current model expands the classical rent-commute logic by viewing relevant geography as an activity system rather than a simple home-to-work trajectory.

### 1.1.2. *Commuting Time, Well-Being, and the Benefit-Costs of Moving*

The literature on commuting time broadens the cost aspect of the model by illustrating that commuting involves more than just vehicle operating costs. The length of a commute influences subjective well-being, life satisfaction, stress levels, health, and the balance between work and personal life. According to Choi et al. (2013), there is a negative correlation between commute duration and both overall subjective well-being and happiness. Similarly, Zijlstra and Verhetsel (2021) demonstrated that the burden of commuting is linked to well-being in various European settings. Reviews on commuting and well-being also concluded that longer commutes often reduce the time available for sleep, leisure, family responsibilities, and social activities (Chatterjee et al., 2020). The reviewed literature on commuting time highlights that acceptable commute durations are frequently discussed concerning practical limits, such as Marchetti's constant (Krasilnikova, 2021), which is typically seen as an approximate one-hour daily travel-time budget, or about thirty minutes each way. Although this benchmark should not be regarded as a universal rule, it is valuable for modeling, as it indicates that households may only endure commuting costs up to a perceived time threshold. When the commute time becomes excessive, relocation, job change, or schedule adjustment becomes more probable. The decision to commute and relocate is influenced by housing affordability. Research on location affordability suggests that housing and transportation costs are inseparable elements of household budgets. The HUD Location Affordability Index and the Center for Neighborhood Technology's Housing and Transportation framework both address the issue that low rent in peripheral areas may be counterbalanced by higher transportation costs (Ganning, 2017; Hamidi et al., 2018; Situ & Kim, 2024). Makarewicz et al. (2020) found that while location-efficient areas can help reduce transportation costs, the financial relief they provide may not be enough for those in the lowest income brackets. Schouten (2021) similarly indicates that transportation savings in dense, transit-rich areas may not fully compensate for high housing costs for low-income households. These findings suggest that affordability cannot be assessed using rent alone. Research that focuses on low-income workers and the issue of spatial mismatch highlights the need to consider constraints. When affordable housing is moved away from areas with abundant job opportunities, low-income workers may experience longer or more difficult commutes, and the imbalance between jobs and housing can lead to increased travel distances (Antipova, 2020; Blumenberg & King, 2019; Blumenberg & King, 2021; Blumenberg & Wander, 2022). Zenou's (2007) search-matching model illustrates how high costs of relocation and labor-market obstacles can exacerbate spatial mismatch. These findings suggest that decisions regarding commuting or relocating are often not free optimization problems, but are instead constrained by factors such as income, housing availability, transportation options, and market barriers.

### 1.1.3. *Socio-Personal Determinants of Utility*

In addition to financial considerations, individual preferences for urban amenities play a crucial role in determining utility. Brueckner, Thisse, and Zenou (1999) illustrated that amenities such as cultural sites and parks can draw affluent individuals to city centers, even when faced with higher expenses or extended travel times, indicating that the benefits of these amenities can surpass the drawbacks of commuting. Social connections also play a vital, yet frequently underestimated, role in residential contentment. Guidon, Wicki, Bernauer, and Axhausen (2019) discovered that being close to social contacts is a key factor in choosing a location, often balanced against the length of the commute. Likewise, Spring, Ackert, Crowder, and South (2017) provided evidence that being near non-coresident relatives discourages moving away and affects destination decisions, as people gain social capital from family networks that keep them tied to certain areas.

Residential mobility is closely connected to life stages. Clark and Huang (2003) demonstrated that life events, such as getting married or having children, influence relocation, shifting the focus to factors like school quality and proximity to family over commuting convenience.

### 1.2. Research Gaps

The extant literature is extensive, yet lacks cohesion. Traditional urban economic models effectively outline the rent-commute trade-off but often oversimplify aspects such as social connections, nonwork accessibility, and institutional context. Although discrete-choice and agent-based models account for diversity and multiple factors, they can become computationally intricate and less clear for practical decision-making. Studies on housing affordability emphasize the 30% income threshold and the combined burden of housing and transportation costs; however, they frequently fail to detail the personal decision-making process regarding whether relocating or commuting is more advantageous. Research on commuting time highlights the negative impact of lengthy commutes on well-being, but it often does not integrate rent comparisons, affordability limits, and social connections into a single algebraic framework. Another gap involves the consideration of indirect and nonwork expenses. Location affordability tools typically merge housing and transportation costs, focusing mainly on work-related travel or general transport expenses. However, the literature, including broader studies on residential choice, indicates that households also travel for medical appointments, family connections, religious activities, education, legal services, and social interactions. These nonwork and relational expenses are crucial to the experience of moving but are seldom explicitly included in simple models comparing moving versus commuting. A further gap pertains to affordability constraints. Although the 30% rule is widely applied, the literature frequently points out its limitations, as it does not account for residual income, transportation, utilities, household composition, or material hardship (Airgood-Obrycki et al., 2022; Shamsuddin & Campbell, 2021). Nevertheless, this threshold remains a useful practical screening tool. A model employing the 30% rule as a feasibility condition while also evaluating the total commute-adjusted costs can maintain the benchmark's practical utility without considering it a comprehensive welfare measure. The final gap is related to interpretability. Many contemporary models use nested logit, hazard models, machine learning, or agent-based simulations to estimate residential choices. While these methods are beneficial, they may be challenging to convey to planners, households, or policy analysts, who require a clear rule. This work, therefore, seeks to develop a straightforward model that is mathematically explicit, theoretically sound, and adaptable.

### 1.3 Significance and Contribution

The proposed model contributes to commuting and residential mobility research in several ways, First, the model is most closely associated with the Alonso-Muth-Wheaton tradition by maintaining the fundamental balance between reduced housing prices and higher costs of commuting (Alonso, 1964; Muth, 1969; Wheaton, 1977). It connects discrete-choice and nested-logit studies by conceptualizing the decision to commute or relocate as a structured choice between remaining and moving, with each option assessed based on cost, accessibility, and household utility (Lee & Waddell, 2010; Mirzahossein et al., 2023). Second, the model also relates to nonwork-accessibility research by considering services and social connections as integral to the household's activity space rather than as external factors (Lee et al., 2010). Additionally, it aligns with location-affordability research by integrating housing and transportation costs and maintaining the 30% affordability threshold as a feasibility constraint rather than a comprehensive measure of well-being (Ganning, 2017; Gabriel & Painter, 2020; Makarewicz et al., 2020; Situ & Kim, 2024). Third, in terms of commuting well-being research, the model includes commuting costs that account for time and welfare losses beyond fuel or vehicle expenses (Chatterjee et al., 2020; Choi et al., 2013; Zijlstra & Verhetsel, 2021). Finally, it is linked to migration and residential mobility research by acknowledging that factors such as social anchoring, risk aversion, neighborhood conditions, and institutional access can explain why households may choose to stay even when moving, which seems financially beneficial (Clark et al., 2017; Graves, 2018; Song & Chung, 2024). This model is well situated within the field of urban economics, particularly within the Alonso-Muth-Mills residential-location tradition, where households balance housing costs against commuting expenses. Additionally, it is relevant to migration and urban geography, where mobility is influenced not only by income and transportation costs but also by social capital, place attachment, family networks, institutions, and subjective well-being.

## 2. Theoretical Model

The following is an algebraic representation of the framework, which has been expanded to incorporate income, affordability constraints, commuting costs, indirect social costs, and place-based benefits. This framework is structured as a two-stage constrained choice model. In the first stage, the model evaluates whether relocation is financially viable and minimizes direct costs. In the second stage, it assesses whether the overall utility of relocating surpasses the utility of remaining in Location A.

### 2.1. *Core variables*

Let the individual currently live in place A and work in place B. Let Ra be the monthly rent in place A, $R_b$ be the monthly rent in place B, I be the monthly income, d be the one-way driving distance from A to the workplace in B, c(d) be the cost of one one-way trip as a function of distance, dc be the daily round-trip driving cost, C(d) be the monthly commuting cost, $W_a$ be the social/place benefit of remaining in A, $W_b$ be the social/place benefit of moving to B, $S_a$ be the indirect social/service-maintenance cost incurred if the person moves to B but continues using services or relationships in A, and $S_b$ be the indirect social/service-access cost incurred if the person remains in A but needs services, amenities, or opportunities in B.

Let $\lambda$ be the weight placed on direct monetary cost, $\theta$ the weight placed on social/place benefits, $\varphi$ the weight placed on indirect social/service costs, and $\alpha$ the affordability threshold set at 0.30.

Assume five workdays per week and roughly four workweeks per month, so the person makes about 20 commuting days per month. With two trips per day:

$$dc = 2c(d)$$
$$C(d) = 20dc = 40c(d)$$

If driving cost is linear in distance, then:

$$c(d) = \kappa d$$

where $\kappa$ is the cost per mile or kilometre. Then:

$$C(d) = 40\kappa d.$$

### 2.2. *Direct-cost model*

The direct monthly cost of remaining in A and commuting to B is:

$$TC_a = R_a + C(d)$$

Given the assumption:

$$TC_a = R_a + 20dc$$

The direct monthly cost of moving to B is:

$$TC_b = R_b$$

Ignoring all non-monetary factors, the individual should move if:

$$TC_a > TC_b$$

Substituting:

$$R_a + C(d) > R_b$$
or:
$$R_a + 20dc > R_b$$

Therefore, the direct-cost decision rule is:

$$M_1 = \begin{cases} 1, & if \text{ Ra} + C(d) > \text{Rb} \\ 0, & if \text{ Ra} + C(d) \leq \text{Rb} \end{cases}$$

where $M_1 = 1$ means move, and $M_1 = 0$ means remain in A and commute.

### 2.3. ***Income and affordability constraints***

Let the rent-affordability threshold be:

$$\alpha I = 0.30I$$

Moving to B is financially feasible only if:

$$R_b \leq 0.30I$$

Remaining in A is financially feasible on rent alone if:

$$R_a \leq 0.30I$$

A stronger affordability condition for remaining in A is
:

$$R_a + C(d) \leq 0.30I$$

The rent-only affordability rules are:

$$\text{A_A\^{}rent} = \begin{cases} 1, & if \text{ Ra} \leq 0.30I \\ & 0, otherwise \end{cases}$$

$$\text{A_B\^{}rent} = \begin{cases} 1, & if \text{ Rb} \leq 0.30I \\ & 0, otherwise \end{cases}$$

The rent-plus-commute affordability rule for staying in A is:

$$A_A^{full} = \begin{cases} 1, \ if \ \text{Ra} + C(d) > \text{Rb} \\ 0, \ if \ \text{Ra} + C(d) \leq \text{Rb} \end{cases}$$

**2.4.** ***Direct-cost decision with income constraint***

A person should consider moving only if moving is both cheaper and affordable:

$$R_A + C(d) > R_b$$
$$R_b \leq 0.30I$$

Thus:

$$M_direct = \begin{cases} 1, \ if \ Ra + C(d) > Rb \ and \ Rb \leq 0.30I \\ 0, otherwise \end{cases}$$

A person should remain in A and commute if:

$$R_a \leq 0.30I$$
$$R_a + C(d) \leq R_b$$

Thus:

$$S_direct = \begin{cases} 1, \text{if Ra} \leq 0.30\text{I and Ra} + \text{C(d)} \leq \text{Rb} \\ 0, otherwise \end{cases}$$

**2.5.** ***Adding indirect social costs and place benefits***

Research on residential mobility, migration, and neighborhood transformation indicates that relocation can lead to social and institutional upheavals. Families may choose to stay put because of the closeness of relatives, reliable service providers, religious centers, familiarity with the neighborhood, continuity in schooling, perceived safety, and informal support systems. Studies on housing vouchers and relocation have revealed that even when families wish for better neighborhood conditions, their options may be limited by factors such as discrimination, credit checks, violence, and stress (Ellen et al., 2022; Graves, 2018; So et al., 2025). Likewise, research on gentrification and displacement has found that while moving can enhance certain neighborhood metrics, it can also heighten stress and uncertainty (Foell et al., 2024; Song & Chung, 2024). These insights support the inclusion of $W_a$, $W_b$, $S_a$, and $S_b$ in the model. $W_a$ signifies the value of social stability in A, whereas $W_b$ denotes the value of opportunities and amenities available in B. $S_a$ reflects the indirect costs of sustaining services and relationships tied to A after relocation, and $S_b$ represents the indirect costs of accessing services in B while remaining in place. This model differentiates between financial feasibility and social appeal. Relocation might be economically sensible but socially burdensome, or it could be financially marginal yet increase utility if B offers significantly better access to jobs, education, healthcare, or other amenities.

The model's direct cost component presupposes that individuals act solely to minimize costs. However, in reality, individuals are also influenced by factors such as social networks, familial connections, attachment to places, institutional affiliations, access to medical services, educational institutions, religious communities, and perceived opportunities. Let Wa denote the social and place-based benefits associated with remaining in location A. These benefits may encompass proximity to family and friends, membership in churches or associations, relationships with trusted professionals such as doctors and lawyers, emotional attachment to the location, familiarity, social identity, and local informal support networks. Conversely, let Wb represent the social, economic, and amenity benefits of relocating to Location B. These benefits may include improved employment opportunities, access to higher-order services, enhanced educational opportunities, urban amenities, professional networking, reduced commuting times, and access to healthcare, cultural, or specialized institutions.

Let $S_a$ be the monthly equivalent indirect cost (C) of moving to B while maintaining important relationships and services in A. For example,

$$S_a = C(doctor_a) + C(friends_a) + C(church_a) + C(legal_a) + C(family_a) + \cdots$$

Let S_B be the monthly-equivalent indirect cost of remaining in A while needing to access opportunities or services in B:

$$S_b = C(services_b) + C(amenities_b) + C(career_b) + C(education_b) + \cdots$$

These costs may be monetary, time-based, psychological, or opportunity-based.

**2.6.** ***Utility formulation***

Utility maximization serves as the primary behavioral framework linking classical urban economics, discrete-choice location models, and agent-based simulations. In models of location choice, it is assumed that households opt for an alternative that maximizes utility, considering constraints related to income, housing market, and transportation. Lee and Waddell (2010) characterize residential location as a combination of housing quality, neighborhood features, and accessibility, whereas agent-based models depict how households choose locations by weighing rent, commuting costs, transport mode, job accessibility, and neighborhood status (Marwal, 2023; Tomasiello et al., 2020). Similarly, Gholami et al. (2024) model agents as reacting to spatial structure through commuting and migration decisions influenced by disposable income after accounting for commuting and housing expenses.

The model presented here adheres to this utility framework but articulates the commute-or-move decision in a concise algebraic form. The utility of remaining in A is expressed as $U_a = \theta W_a - \lambda[R_a + C(d)] - \varphi S_b$, whereas the utility of relocating to B is given by $U_b = \theta W_b - \lambda R_b - \varphi S_a$. This formulation integrates monetary costs, social benefits, and indirect access costs into a unified decision rule while maintaining distinct weights for direct cost sensitivity, place utility, and social service maintenance burdens. The resulting condition for moving, $\Delta U > 0$, identifies the point at which the weighted utility of relocating surpasses that of the remaining. Behavioral research also indicates that residential choices are influenced by factors beyond objective costs, such as risk aversion, attachment, tenure, and familiarity. Research grounded in prospect theory on migration suggests that risk aversion and the endowment effect can heighten the likelihood of staying, as familiarity, social connections, and perceived security in the current location enhance the subjective value of the remaining (Clark et al., 2017). This study theoretically supports $W_a$ as a social anchoring term and implies that the model should not automatically favor moving whenever direct costs suggest relocation; a significant $W_a$ or $S_a$ may justifiably counterbalance the financial savings from moving.

Define the total utility of staying in A as:

$$U_a = \theta W_a - \lambda[R_a + C(d)] - \varphi S_b$$

Define the total utility of moving to B as:

$$U_b = \theta W_b - \lambda R_b - \varphi S_a$$

The person should move if:

$$U_b > U_a$$

Substituting:

$$\theta W_b - \lambda R_b - \varphi S_a > \theta W_a - \lambda[R_a + C(d)] - \varphi S_b$$

Rearranging:

$$\lambda[R_a + C(d) - R_b] + \theta(W_b - W_a) + \varphi(S_b - S_a) > 0$$

Therefore, define the net move advantage:

$$\Delta U = U_b - U_a$$
$$\Delta U = \lambda[R_A + C(d) - R_B] + \theta(W_B - W_A) + \varphi(S_B - S_A)$$

The general move rule is:

$$M = \begin{cases} 1, if\ \Delta U > 0\ and\ Rb \leq 0.30I \\ 0, otherwise \end{cases}$$

That is:

$$M = \begin{cases} 1, if\ \Delta U > 0\ and\ Rb \leq 0.30I1,\ if\ \lambda[Ra + C(d) - Rb] + \theta(Wb - Wa) + \varphi(Sb - Sa) > 0\ and\ Rb \leq \\ 0, otherwise \end{cases}$$

## 2.7. *Interpretation of the model*

The term $R_a + C(d) - R_b$ is the direct financial saving from moving. If it is positive, then moving to B is cheaper than living in A and commuting. The term $W_b - W_a$ is the net place-benefit advantage of B. If positive, B offers greater social, amenity, institutional, or opportunity value than A. The term $S_b - S_a$ is the net indirect-cost advantage of moving. If positive, then staying in A creates more indirect access costs than moving to B. If negative, then moving creates greater indirect costs because the person remains socially or institutionally anchored in A.

### 2.8. Mathematical proofs

#### 2.8.1. *Proof 1 - Direct-cost threshold*

The person compares:

$$TC_a = R_a + C(d)$$
$$TC_b = R_b$$

Moving is preferred when:

$$TC_b < TC_a$$

Substituting:

$$R_b < R_a + C(d)$$

which is equivalent to:

$$R_a + C(d) - R_b > 0$$

Therefore, moving is directly cost-minimizing if and only if:

$$R_a + C(d) > R_b.$$

This proves the first decision rule.

#### 2.8.2. *Proof 2 - Distance threshold for moving*

Assume commuting cost is linear:

$$C(d) = 40\kappa d$$

The direct-cost move condition is:

$$R_a + 40\kappa d > R_b$$

Subtract R_A from both sides:

$$40\kappa d > R_b - R_a$$

Divide by 40κ, assuming $\kappa > 0$:

$$d > (R_b - R_a) / 40\kappa$$

Define the critical commuting distance:

$$d^* = (R_b - R_a) / 40\kappa$$

Thus:

$$M_direct = 1 \Leftrightarrow d > d^*$$

If $R_b \leq R_a$, then $R_b – R_a \leq 0$, so $d^* \leq 0$. Since distance is nonnegative, $d \geq 0$, moving is direct-cost preferred for any positive commute distance when B has rent equal to or lower than A.

### 2.8.3. *Proof 3 - Income-constrained feasibility*

The affordability constraint for moving is:

$$R_b \leq 0.30I$$

The direct-cost condition is:

$$R_a + C(d) > R_b$$

Both must hold for a financially feasible move:

$$\{R_a + C(d) > R_a\} \cap \{R_b \leq 0.30I\}$$

Therefore:

$$M_direct = 1 \Leftrightarrow R_a + C(d) > R_b \text{ and } R_b \leq 0.30I$$

If either condition fails, moving is not recommended by the direct financial rule. Specifically, if $R_a + C(d) \leq R_b$, moving is not cheaper. If $R_b > 0.30I$, moving may be cheaper than commuting but still unaffordable.

### 2.8.4. *Proof 4 - Utility-based move condition*

The person moves if:

$$U_b > U_a$$

Then:

$$U_b - U_a > 0$$

Substitute:

$$[\theta W_b – \lambda R_b – \varphi S_a] - [\theta W_a - \lambda(R_a + C(d)) – \varphi S_b] > 0$$

Expand:

$$\theta W_b – \lambda R_b – \varphi S_a – \theta W_a + \lambda(R_a + C(d)) + \varphi S_b > 0$$

Collect terms:

$$\lambda[R_a + C(d) – R_b] + \theta(W_b – W_a) + \varphi(S_b – S_a) > 0$$

Therefore:

$$M = 1 \Leftrightarrow \Delta U > 0$$

where:

$$\Delta U = \lambda[R_a + C(d) - R_b] + \theta(W_b - W_a) + \varphi(S_b - S_a)$$

### 2.9. *Full decision rule*

The full decision rule can be expressed as:

$$M = \begin{cases} 1, if\ Rb \leq 0.30I \text{ and } \Delta U > 0 \\ 0, otherwise \end{cases}$$

where:

$$\Delta U = \lambda[R_a + C(d) - R_b] + \theta(W_b - W_a) + \varphi(S_b - Sa).$$

A stricter version, closer to the original two-rule logic, is:

$$M = \begin{cases} 1, if\ Ra + C(d) > Rb, Rb \leq 0.30I, and\ Wb > Wa \\ 0, otherwise \end{cases}$$

This decision rule says that moving is recommended only when moving reduces direct monthly cost, rent in B is affordable, and B has greater social/place benefit than A. However, the weighted utility version is more flexible because it allows a very large benefit in B to compensate for modestly higher rent, or a very strong attachment to A to prevent moving even if B is cheaper.

### 2.10. *Suggested final decision rule*

A more robust expression of the final decision rule:

$$M = \begin{cases} 1, if\ Rb \leq 0.30I\ and\ \lambda[Ra + C(d) - Rb] + \theta(Wb - Wa) + \varphi(Sb - Sa) > 0 \\ 0, otherwise \end{cases}$$

with:

$$C(d) = 20dc$$
$$dc = 2c(d)$$
$$C(d) = 40c(d)$$

If $c(d) = \kappa d$, then:

$$C(d) = 40\kappa d$$

Therefore:

$$M = \begin{cases} 1, if\ Rb \leq 0.30I\ and\ \lambda[Ra + C(d) - Rb] + \theta(Wb - Wa) + \varphi(Sb - Sa) > 0 \\ 0, otherwise \end{cases}$$

This constitutes the algebraic core of the proposed model.

## 3. Discussion

The decision to either commute or relocate is a complex issue that is most effectively analyzed through utility modeling, which considers a wide range of factors beyond mere economic costs. Key determinants of an individual's perceived utility include personal preferences such as access to amenities, the strength of social networks, and proximity to family and essential services. Gaps in research complicate modeling since personal preferences are incredibly complex, and unique to each household situation. The current model addresses these gaps by integrating direct costs, affordability constraints, social benefits, and indirect access costs into an algebraic framework. The direct-cost rule, $R_a + C(d) > R_b$, translates the classical bid-rent and housing-transportation trade-off into a two-location decision rule. The income constraint $R_b \leq 0.30I$ is grounded in the common housing affordability standard that housing expenditures should remain below 30 percent of earnings. This benchmark is commonly applied in both policy and research contexts, yet it faces criticism for potentially overlooking factors such as remaining income, family size, essential non-housing expenses, and regional cost variations. The stay condition, $R_a \leq 0.30I$ and $R_a + C(d) \leq R_b$, recognizes that remaining in A may be rational when rent is affordable and commute-adjusted costs remain below the rent in B.

The model's direct-cost component is closely aligned with the Alonso-Muth-Mills monocentric city model ,wherein households choose residential locations by balancing land or housing costs against commuting costs. Locations farther from the employment center tend to have lower rents but higher commuting costs. Locations closer to work tend to have higher rents but lower commuting costs. The core tradeoff in this model, R_A + C(d) versus R_B, is essentially a simplified two-location version of this rent-commute tradeoff. Muth's (1969) residential location theory and later urban economic work emphasize that income, housing demand, and access to employment jointly influence where households locate. Diamond's (1980) revisit of Muth's (1969) framework shows that income and residential location are linked by a desire for housing space and access. The addition of an income constraint is therefore theoretically appropriate because affordability is not merely a comparison between two rents; it depends on whether either rent is feasible, given household income.

The commuting-cost function C(d) also reflects a long-standing urban geography assumption: distance imposes both monetary and time costs. Transport improvements, driving costs, and speed affect the attractiveness of residential locations because they change the effective cost of distance. More recent quantitative urban economics continue to treat commuting friction as central to urban structure and residential sorting. Another strength of the model is that commuting was not treated as a neutral transfer of money. Commutes impose fatigue, stress, time loss, health burden, and reduced subjective well-being. Thus, the commuting cost function can possibly be expanded to account for these factors at the personal level in the modeling process.

The model improves on the simplest monocentric framework by recognizing that people not only commute to work. They also travel for healthcare, education, worship, friendship, legal services, childcare, shopping, recreation, and family obligations. This aligns with research showing that nonwork accessibility matters for residential location. Lee, Waddell, Wang, and Pendyala (2010) argue that residential choice should incorporate both work and nonwork accessibility, because households evaluate locations based on a broader daily activity space, not merely the workplace commute. Migration research has long shown that relocation is socially embedded. People often remain in places where they have family, friends, trusted institutions, and emotional attachment, even when another location appears to be economically superior.

Therefore, $W_a$ and $W_b$ were important. They allow the model to represent place utility, that is, the subjective value of a location. The model also addresses the gap between work-commute and broader activity-space models. By including $S_a$ and $S_b$, households maintain social, medical, religious, educational, and professional relationships across spaces. This links the model to Lee et al.'s (2010) finding that nonwork accessibility matters for residential location choice and to migration research emphasizing place attachment, risk aversion, and the endowment effect (Clark et al., 2017). In this sense, the model treats moving not merely as a change in address, but as a reorganization of an individual's social and service geography.

Finally, the model's utility extension links directly to discrete choice and agent-based approaches by representing the household as a utility-maximizing actor. However, rather than estimating a full multinomial or nested logit model, it expresses the decision as $\Delta U = \lambda[R_a + C(d) - R_b] + \theta(W_b - W_a) + \varphi(S_b - S_a)$. This formulation is transparent and interpretable. The term $R_a + C(d) - R_b$ captures the direct financial advantage of moving, $W_b - W_a$ captures the net place-benefit advantage of B, and $S_b - S_a$ captures whether the indirect access burden is greater if the household stays or moves. The move rule $M = 1$ if $\Delta U > 0$ and $R_b \leq 0.30I$, therefore, preserves the logic of utility maximization while making the decision rule operational.

## 4. Conclusion

The choice between commuting and relocating involves a complex set of factors that go beyond the simple comparison of housing and commuting costs. While traditional urban economic theories offer a foundational perspective, recent studies reveal that factors such as affordability limits, transportation expenses, job accessibility, social connections, emotional ties to a place, institutional access, and personal well-being also influence residential mobility. This paper adds to the existing body of work by presenting a mathematically detailed framework that incorporates these elements into a unified utility-based decision model. By integrating direct housing and commuting expenses with income limitations, indirect costs related to social and service access, and location-based utility, the model offers a clear and understandable depiction of the decision to commute or move. Unlike more complex residential location models, this framework provides a straightforward algebraic decision rule that is easily comprehensible, adaptable, and expandable for researchers, planners, and policymakers. The model reveals that financially optimal choices do not always align with maximizing household utility, as factors like social ties, service access, and neighborhood features might justify staying put even when moving seems economically beneficial. Similarly, a more expensive destination might offer enough benefits in terms of accessibility, job prospects, or quality of life to warrant relocation despite higher housing costs. Overall, the framework underscores the need to assess housing affordability and transportation expenses together rather than separately. By integrating economic, geographic, and behavioral factors into a single mathematical model, this research lays the groundwork for future empirical validation and adjustment using household survey or administrative data. Consequently, the model serves as both a theoretical advancement in urban economics and residential mobility studies and a practical tool for assessing relocation and commuting choices in increasingly intricate metropolitan housing and labor markets.

## 5. Practical implications

The proposed model holds significant practical value for both residential mobility modeling and public policy. In terms of modeling, it offers a clear and mathematically defined framework that connects traditional urban economic theory with more complex computational methods like discrete-choice, nested-logit, and agent-based models. By incorporating factors such as housing costs, commuting expenses, income-related affordability constraints, indirect social costs, and place-based utility into a single algebraic equation, the model presents an understandable decision-making rule that can be easily tailored to various metropolitan regions, household profiles, transportation options, or policy scenarios. This adaptability provides researchers with a versatile tool for analyzing residential location choices, while also being accessible to planners, policymakers, and practitioners who may not need highly detailed simulation models.

From a policy standpoint, the model emphasizes the necessity of considering housing affordability and transportation costs together rather than as separate policy areas. The findings indicate that assessing residential affordability solely through rent or mortgage payments is insufficient, as transportation costs and commuting time can significantly impact the actual financial burden on households. Thus, the framework advocates for integrated housing and transportation planning by demonstrating how investments in transit infrastructure, location-efficient affordable housing, employment decentralization, and mixed-use development can affect relocation and commuting choices. Furthermore, since the model explicitly accounts for social attachment, service access, and neighborhood utility, it acknowledges that financially optimal decisions may not always enhance household welfare. As a result, the framework offers policymakers a more comprehensive foundation for assessing relocation incentives, housing assistance programs, and urban development strategies aimed at improving both economic efficiency and household well-being.

## 6. Directions for future research

Future research should aim to enhance utility models to more thoroughly capture the intricate interdependencies involved, thereby providing a more detailed understanding of urban migration and journey-to-work patterns. Cost decision models need refinement to incorporate housing, transportation, utilities, and other essential expenditures, while considering the diversity of household preferences and constraints related to income, race, and life-cycle stages. Policy interventions must address both supply-side constraints, such as increasing affordable housing in areas with abundant job opportunities, and demand-side support, including expanded rental assistance and reformed screening practices, to facilitate genuine choice in the commute-or-move decision for economically vulnerable households.

## 7. Limitations

There are limitations to the model. First, the model assumes a single workplace. In reality, people may change jobs, work remotely, have hybrid schedules, or belong to households with multiple workers. For example, a two-worker household would require two commuting-cost functions, and a hybrid worker would require fewer commute days. Second, although the model treats income as a fixed variable, moving to B may increase job opportunities and expected wages. Therefore, a more advanced model would use expected income. Third, the model does not include moving or utility costs. Fourth, housing quality could be added using some weight variable to denote the perceived quality of housing in places A and B. Finally, the model does not include household composition. Children, elderly dependents, school districts, caregiving obligations, and partner employment can dominate this decision. These could be incorporated in $W_a$, $W_b$, $S_a$, and $S_b$; however, a more explicit household model would be preferable.

## 8. Funding

This research did not receive any specific grants from funding agencies in the public, commercial, or not-for-profit sectors.

## 9. Conflict of interest

The authors declare no conflicts of interest.